\documentclass{aa}

\usepackage{graphicx}
\usepackage{txfonts}
\usepackage{lipsum}
\usepackage{subcaption}         
\usepackage{lscape}             
\usepackage{placeins}           

\begin{document}

   \title{Study of photometric and spectral variability of the roAp star HD~210684}


   \author{V. Khalack\inst{1}
        \and C. Lovekin\inst{2}
        }

   \institute{D\'epartement de Physique et d'Astronomie, Universit\'e de Moncton, Moncton, N.B., Canada E1A 3E9\\
             \email{Viktor.Khalack@umoncton.ca}
            \and Department of Physics, Mount Allison University, Sackville, N.B., Canada E4L 1E6\\ }

   \date{Received January 13, 2026}


  \abstract
{}
   {
This paper studies photometric and spectropolarimetric variability of HD~210684 in order to derive its magnetic properties, rotational period, evolutionary stage, and global stellar parameters. }
   {
The Discrete Fourier Transform is used to measure frequencies and amplitudes 
of periodic signals present in light curves of HD~210684. Evolution models are calculated with MESA, 
while roAp type pulsations are simulated with GYRE. The values of $T_{\rm eff}$, $\log(g)$, v$\sin{i}$, and radial velocity (RV) are derived from the best fit of Balmer line profiles using FITSB2. The Least Square Deconvolution (LSD) method is applied to available Stokes I \& V spectra to measure the mean longitudinal magnetic field  $\langle B_{\rm z}\rangle$
and RV. }
   { Detailed analysis of photometric variability of the light curves provided by \textit{TESS} for HD~210684 reveals rotational modulation with period P=5.02188$\pm$0.00005~d and splitting of high-overtone pulsations that corresponds to the same rotational period. 
Derived $\langle B_{\rm z}\rangle$ measurements also show periodic variability with P=5.02188~d, but this variability appears to be different from the one predicted by a centred magnetic dipole model, suggesting that the configuration of the surface magnetic field is more complex than the dipolar one. We have determined the inclination angle between the line of sight and the rotation axis to be $i = 31\degr \pm 2\degr$, and the angle between the rotation and magnetic dipole axes  as $\beta= 77\degr \pm 3\degr$. Simulations of stellar pulsations constrain $\log T_{\rm eff} < 3.85$ and show that our best fitting model depends on whether the observed modes are $\ell = 1$ or $\ell =2$.  }
  {Considering the derived value of $\beta$, we prefer the best fit model with $\ell =2$ mode, which predicts that HD~210684 lies on the main sequence with an age of approximately 1.45~Gyr. Meanwhile, only the most evolved and coolest models in our grid show $\eta > 0$ (when frequencies are driven) for the observed triplet. The best fit of Balmer line profiles has resulted in higher values of $T_{\rm eff}$ and $\log(g)$ for spectra acquired at rotational phases $\varphi$= 0.14 -- 0.39 suggesting visibility of an area with higher surface temperature. Strong asymmetry of the LSD Stokes I profiles derived at $\varphi$= 0.307 and 0.394 supports this hypothesis. Clear variability of the LSD Stokes I profiles with rotational phase argues for horizontal abundance stratification that may be linked to the inhomogeneity of surface temperature. Further spectropolarimetric observations of HD~210684 are required to study its unique properties in detail.
}

\maketitle
\nolinenumbers
\section{Introduction}
\label{Intro}

The Transiting Exoplanet Survey Satellite (\textit{TESS}) \citep{Ricker+15} launched in April of 2018, has collected a huge amount of high-quality photometric data
for stellar targets located over the whole sky during the last 7 years of its operation. Extensive analysis of the \textit{TESS} data provided by Mikulski Archive for Space Telescopes\footnote{https://mast.stsci.edu/portal/Mashup/Clients/Mast/Portal.html} 
(MAST) allowed to discover new candidate roAp stars \citep[see for example][]{Mathys+23} and confirm the presence of high-overtone pulsations in the known roAp stars located in Southern (using the cycle-1 data) \citep{Holdsworth+21} and Northern (using the cycle-2 data) \citep{Holdsworth+24} hemispheres. HD~210684 was first identified as a roAp type candidate by O.~Kobzar in 2020 with rotational modulation (P = 5.04$\pm$0.02~d) during the analysis of the cycle-2 data for his master's thesis.

HD~210684 (HIP~109552, TIC~259017938) is located 142.71$\pm$0.46~pc from our Sun \citep{Brandt21} and according to \citet{Cannon+Pickering+93} has been assigned {\bf the F0} spectral type. By studying its long-term proper motion in the Hipparcos data and Gaia’s second data release (GDR2) catalogs \citet{Kervella+19} have detected a proper motion anomaly consistent with the presence of a perturbing secondary object of mass M=79.35$\pm$9.63 M$_J$ located at distance around 1.89 A.U. from the primary.
\citet{Balona22} has reported a detection of rotational modulation with period P = 5.102~d in the \textit{TESS} light curve of HD~210684 
and marked it as a probable roAp candidate. Based on the clear detection of the high-overtone pulsations at 1.352116~mHz, \citet{Holdsworth+24} have confirmed its roAp nature.

The roAp stars are known to possess chemically peculiar stellar atmospheres \citep{Gelbmann98, Ghazaryan+19} where convection is suppressed by a relatively strong magnetic field and atomic diffusion can lead to horizontal and vertical abundance stratification \citep{Michaud70, Alecian+Stift21, Alecian23}.
Since 2020, when HD210684 was still a roAp candidate, regular spectropolarimetric observations of this target have been carried out at the Canada-France-Hawaii Telescope (CFHT) to accumulate Stokes I \& V spectra suitable for abundance analysis and for measurements of magnetic field, which is expected to be found in roAp stars.
Detection of a significant mean longitudinal magnetic field in HD~210684 and its variability with the period of stellar rotation are crucial additional {\bf evidence} of its roAp nature. Expected variability of Stokes I \& V line profiles with the same rotational period will indicate presence of peculiar abundance patches in stellar atmosphere of the studied target, and open {\bf the} door for abundance mapping and reconstruction of magnetic field configuration.

Therefore, in this article, we aim to carry out an extensive analysis of the available data for the recently discovered roAp variable HD~210684. We combine its \textit{TESS} photometric data with high-resolution Stokes I \& V spectra obtained with the spectropolarimeter ESPaDOnS to study the pulsation properties of this target, derive its rotation period and global stellar parameters,
measure the mean longitudinal magnetic field and the period of its variability.
Photometric and spectropolarimetric observations of HD~210684 are described in Section~\ref{Observations}. Results of our analysis 
and lists of the derived global stellar parameters 
are presented in Section~\ref{analysis}. Discussion follows in Section~\ref{discussion}.

\section{Observations}
\label{Observations}

\subsection{\textit{TESS} photometry and data reduction}
\label{photo}

HD~210684 (TIC~259017938) has been observed by \textit{TESS} 
\citep{Ricker+15} several times during cycles 2, 5 and 7 of its operation (see Table~\ref{tab_tess}).
During each cycle, \textit{TESS} covers approximately half of the sky, divided into 13 sectors. 
Each sector is monitored for $\sim~27.4$~d using four wide-field cameras covering $24\degr\times24\degr$ each. Data for each sector includes a few gaps; at least one for telescope rotation, sometimes, multiple due to bad data.



\begin{table}
\begin{center}
\caption{Journal of photometric observations of HD~210684 (TIC~259017938) with \textit{TESS}. }
\label{tab_tess}
\begin{tabular}{clrcc}\hline
Sector & BJD (start)& $t_{\rm exp}$ & Background & Data \\
       &(2400000+)& (s)  & subtraction& type\\ 
\hline
15 & 58711.389832 & 1426 & sky background & \textit{TESS}cut \\
15 & 58711.389832 &  120 &                & \textit{TESS}-SPOC \\
16 & 58738.681763 & 1426 & PCA detrended  & \textit{TESS}cut \\
16 & 58738.681763 &  120 &                & \textit{TESS}-SPOC \\
56 & 59825.262653 &  158 & sky background & \textit{TESS}cut \\
83 & 60559.443820 &  158 & sky background & \textit{TESS}cut \\
\hline
\end{tabular}
\end{center}
\end{table}

The reduction procedure of the 24$\times$24 pixel images of the sky-area around HD~210684 is described in detail by \citet{Labadie-Bartz+22, Labadie-Bartz+23}.
To remove the sky background and instrumental trends, the Principal Component Analysis (PCA) detrending method with five regressors was used for the data observed with 1426~s cadence in sector 16 (see Table~\ref{tab_tess}).
In the case 
where the PCA detrending method results in some flux perturbations in the vicinity of observation gaps, 
a simple sky background subtraction method was used to extract the reduced light curve (LC) in sectors 15, 56 and 83. 
Then, the reduced LCs were cleaned of outliers showing more than 4-sigma deviations (where the mean value and standard deviation were calculated for a sliding time window of a given width) and of photometric measurements strongly contaminated by instrumental noise (usually at the beginning or/and at the end of \textit{TESS}' observation in each sector). To ensure a proper evaluation of the mean value and standard deviation for the sliding time window the width of 71300~sec. was used to remove outliers from data in sectors 15 \& 16, and of width 31600~sec for data in sectors 56 \& 83.
The reduced flux and {\bf its} 
error-bars were transformed {\bf into} stellar magnitudes for further analysis.

To study the splitting of high-overtone pulsations (see Subsection~\ref{variability}) we have also used light curves observed with 120~seconds cadence from sectors 15 \& 16 (see Table~\ref{tab_tess}) and produced by the \textit{TESS} Science Processing Operations Center (SPOC) pipeline\footnote{https://archive.stsci.edu/hlsp/tess-spoc} \citep{Caldwell+20}.

\subsection{Spectropolarimetry with ESPaDOnS}
\label{spectropol}

\begin{table}
\begin{center}
\caption{Journal of spectropolarimetric observations with ESPaDOnS. }
\label{tab_spectra}
\begin{tabular}{llrc}\hline
 Name & HJD & $t_{\rm exp}$ & SNR   \\ 
      &(2400000+)& (s)  &  Stokes I/V  \\
\hline
 2555627 & 59177.70782 & 880 & 90/80  \\
 2556314 & 59180.71352 & 880 & 420/370 \\
 2634276 & 59459.89670 & 640 & 400/360 \\
 2690049 & 59542.68543 & 640 & 410/370 \\
 2690811 & 59545.68669 & 640 & 400/340 \\
 2691267 & 59547.68653 & 640 & 240/210 \\
 2776992 & 59805.12729 & 880 & 500/460 \\
 2782172 & 59832.00028 & 880 & 500/450 \\
 2782907 & 59835.83241 & 880 & 530/480 \\
 2783031 & 59836.97868 & 880 & 520/450 \\
 2783328 & 59837.95679 & 880 & 510/470 \\
 2788523 & 59868.84363 & 880 & 520/470 \\
 2875391 & 60093.10957 & 880 & 320/290 \\
 2918356 & 60237.84741 & 880 & 410/320 \\
 2918718 & 60238.86927 & 880 & 460/340 \\
 2919861 & 60243.78181 & 880 & 490/360 \\
 2942163 & 60280.71959 & 880 & 480/430 \\
 2942721 & 60282.77044 & 880 & 480/430 \\
 2950026 & 60316.68966 & 880 & 480/420 \\
 3169052 & 60773.14071 & 880 & 430/380 \\
 3210624 & 60896.87912 & 880 & 410/360 \\
 3252228 & 61017.75834 & 880 & 430/400 \\
\hline
\end{tabular}
\end{center}
\end{table}

After detection of rotational modulation in the photometric data accumulated by \textit{TESS} for HD~210684 during cycle 2 (sectors 15 \& 16) this roAp star was selected for follow-up spectropolarimetric observations with the spectropolarimeter ESPaDOnS (Echelle SpectroPolarimetric Device for Observations of Stars)\footnote{CFHT's web page for ESPaDOnS: https://www.cfht.hawaii.edu/Instruments/Spectroscopy/Espadons/} installed at CFHT. 
ESPaDOnS uses the deep-depletion e2v device Olapa that allows it to produce high-resolution (R=65000) Stokes I, V, Q \& U spectra in the spectral domain from 3700\AA\, to 10000\AA. The optical characteristics of the instrument and its performance are described in detail by \citet{Donati+06}\footnote{For more details about this instrument, the reader is invited to visit {\rm www.cfht.hawaii.edu/Instruments/Spectroscopy/Espadons/}}. The dedicated package Libre-ESpRIT \citep{Donati+97} was employed by the CFHT's team to reduce the Stokes I spectra and the Stokes V circular polarisation obtained for HD~210684.

This target has been observed with ESPaDOnS during several semesters starting from the 2021B semester (CFHT) and finishing with the 2025B semester (CFHT). Luckily for us, the acquired high-resolution and high signal-to-noise ratio (SNR) Stokes I \& V spectra cover the time interval before and after the time-interval of \textit{TESS} observation of HD~210684 in sectors 56 and 83 (see Tables~\ref{tab_tess}, \ref{tab_spectra}). This fortunate time coverage allows us to measure the values of rotational phases with high precision, as is required for
a study of the mean longitudinal magnetic field variability.
The first spectrum of the target acquired on HJD= 2459177.70782 has a relatively small SNR (<100, see 4$^{th}$ column in the Table~\ref{tab_spectra}) for Stokes I \& V spectra at the echelle order \#38 centered at 
596~nm 
due to bad weather conditions. 
Considering that all spectropolarimetric observations of HD~210684 were acquired by ESPaDOnS in the \textit{snapshot mode}, we decided to keep a relatively high exposure time and requested a high SNR {\bf of over 400 per four exposures} 
even under poor weather conditions.

\section{Data analysis}
\label{analysis}

\begin{figure}
\begin{center}
	\includegraphics[width=\columnwidth,angle=-90, scale=0.65]{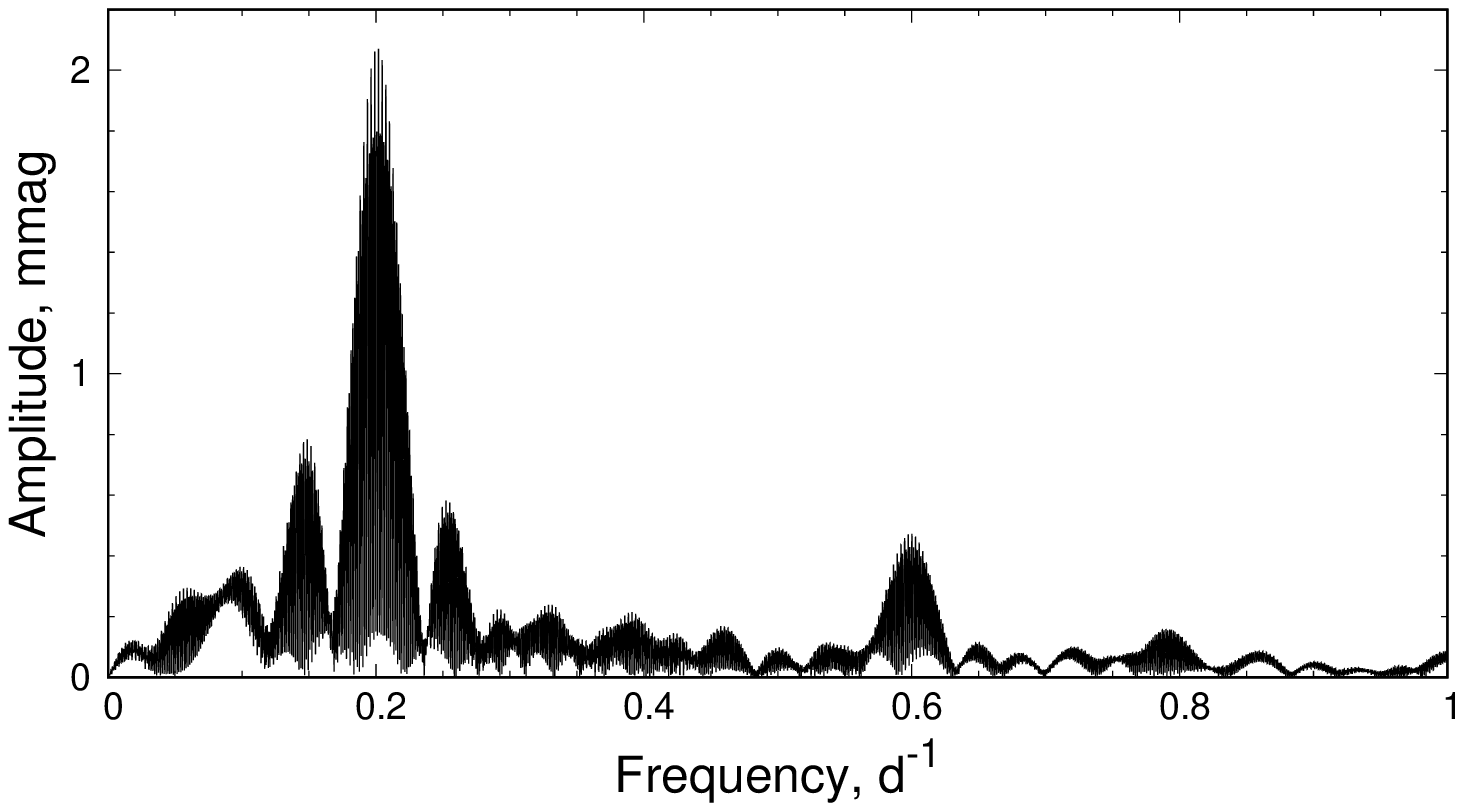}
\end{center}
\caption{DFT of the combined photometric data obtained for the low frequency domain. The most prominent signal and its overtones are caused by rotational modulation of the studied LC. }
\label{fig:rot_modulation}
\end{figure}

\begin{figure}
\begin{center}
	\includegraphics[width=\columnwidth, angle=-90, scale=0.65]{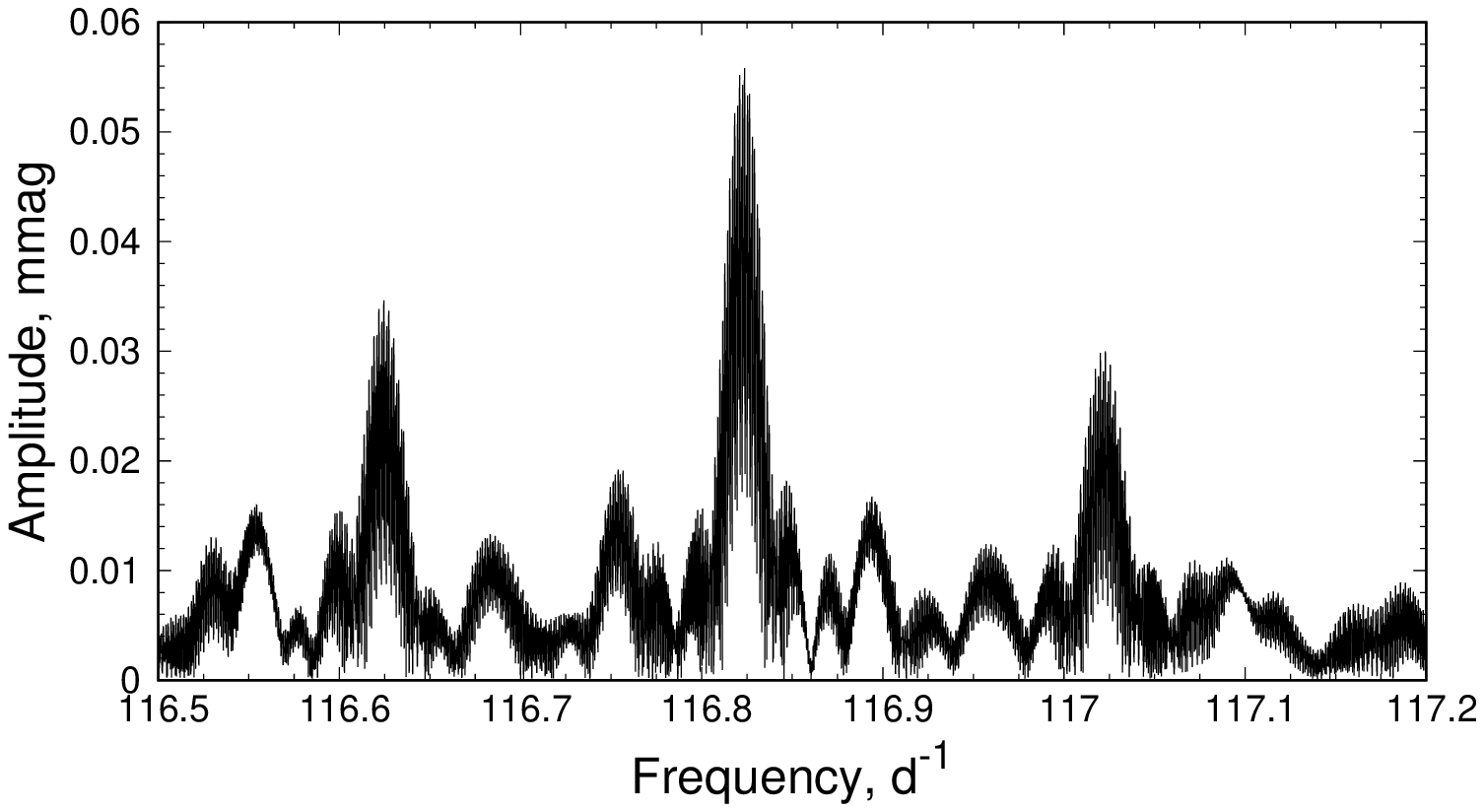}
\end{center}
\caption{The same as Figure~\ref{fig:rot_modulation}, but for a higher frequency range. Frequencies of the three detected signals 
are split with the rotational frequency which allowed to confirm our correct 
estimation of the rotational period. }
\label{fig:puls}
\end{figure}

\begin{figure}
\begin{center}
\includegraphics[width=\columnwidth]{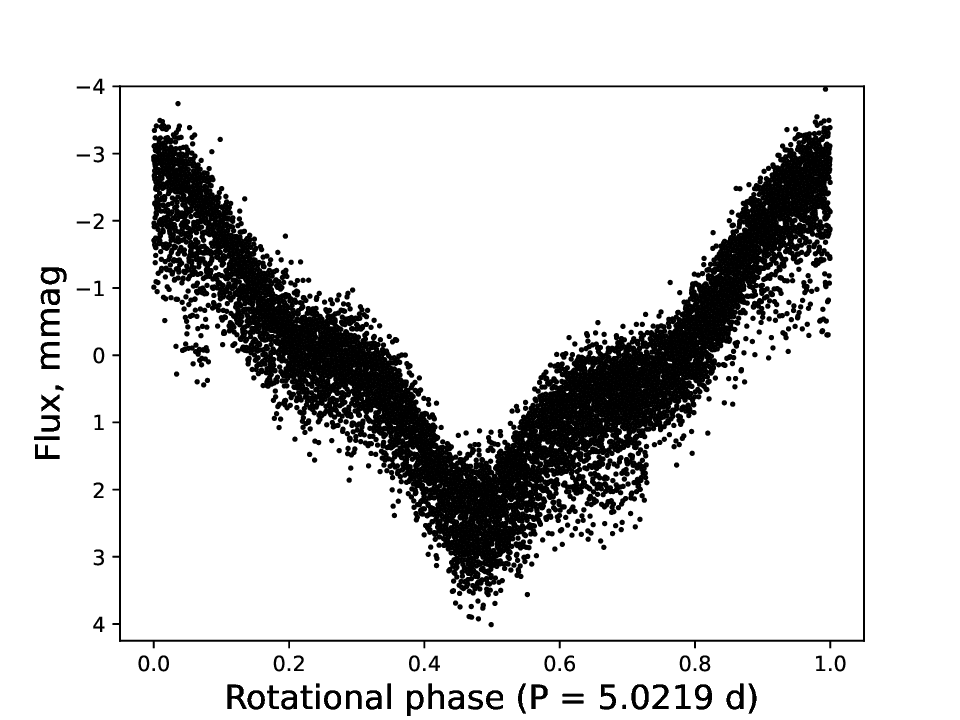}
\end{center}
\caption{Light curve of HD~210684 folded with the derived rotational period.
}
    \label{fig:phased_LC}
\end{figure}

\subsection{Study of photometric variability of HD~210684}
\label{variability}

The reduced data on \textit{TESS} stellar photometry from all four available sectors were combined and analysed with the help of \textit{Period04} software \citep{Lenz+Breger05}. We have employed the Discrete Fourier Transformation (DFT) method to derive frequencies with significant (SNR>4) amplitudes at the low frequency domain (see Fig.~\ref{fig:rot_modulation}). Error bars for the resulting frequencies, amplitudes and phases were calculated using the least-square method and Monte-Carlo method, both implemented in \textit{Period04}. The most significant peak appears at the frequency $\nu_{\rm rot}$=0.1991286$\pm$0.000002\,d$^{-1}$ that corresponds to the rotation period P=5.02188$\pm$0.00005~d.
This period is used to estimate rotational phases for each obtained spectrum (see Table~\ref{tab_Balmer}).

We have applied the code \textit{Period04} to \textit{TESS}-SPOC 120~sec. cadence data from sectors 15 \& 16 combined with \textit{TESS}cut 158~sec. cadence data from sectors 56 \& 83 (see Table~\ref{tab_tess}), and have
{\bf found} three significant signals (see Fig.~\ref{fig:puls}) that correspond to the roAp type stellar pulsations \citep[for more details, see][]{Kurtz82}. The three detected signals are split by the frequency $\nu_{\rm split}$= 
0.19913$\pm$0.00005\,d$^{-1}$ which appears to be the same as the frequency of stellar rotation. 
The \textit{curve\_fit}-routine provides a phase and 
BJD of the first maximum observed in sector 56, and their error-bars. The phased LC is shifted by phase such that the zero-phase corresponds to the maximum detected flux (see Fig.~\ref{fig:phased_LC}). This approach allows us to derive the following ephemerides 
for the rotational modulation in HD~210684 :
\[E = 2459829.53514\pm0.00001+5.02188\pm0.00005 {\rm \,\,BJD,} \]
where the first member specifies BJD for the 
LC maximum, 
while the second member provides rotational period.
Figure~\ref{fig:phased_LC} shows
non-sinusoidal variability with the flux minimum located close to the phase $\varphi$=0.45, and some "flat" areas around $\varphi$=0.25 and 0.7. 

\subsection{Rotational splitting}
\label{puls_results}

To fit the detected frequencies assuming their rotational splitting we have employed MESA version 12778 \citep{Paxton+11, Paxton+13, Paxton+15, Paxton+18, Paxton+19} to calculate a grid of stellar structure and evolution models, and GYRE \citep{Townsend+Teitler13, Townsend+18} to simulate linear non-adiabatic pulsations for each derived model (see Appendix~\ref{sim_pulsation} for more details).

With only one observed frequency triplet, we do not expect to be able to tightly constrain the star based on comparison to the theoretical models, although we are able to rule out some regions of the parameter space. To determine which models fit the observations, we identified a subset of 223 models where the central frequency of the model triplet was within 0.05~d$^{-1}$ of the observed value. We then determined which of these models showed the lowest difference from the observed splitting. This process was done separately assuming the triplet was an $\ell = 1$ and {\bf an} $\ell = 2$ mode. The parameters for the two best fitting models are summarized in Table~\ref{tab:fitresults}.

\begin{figure}
    \includegraphics[width=0.5\textwidth]{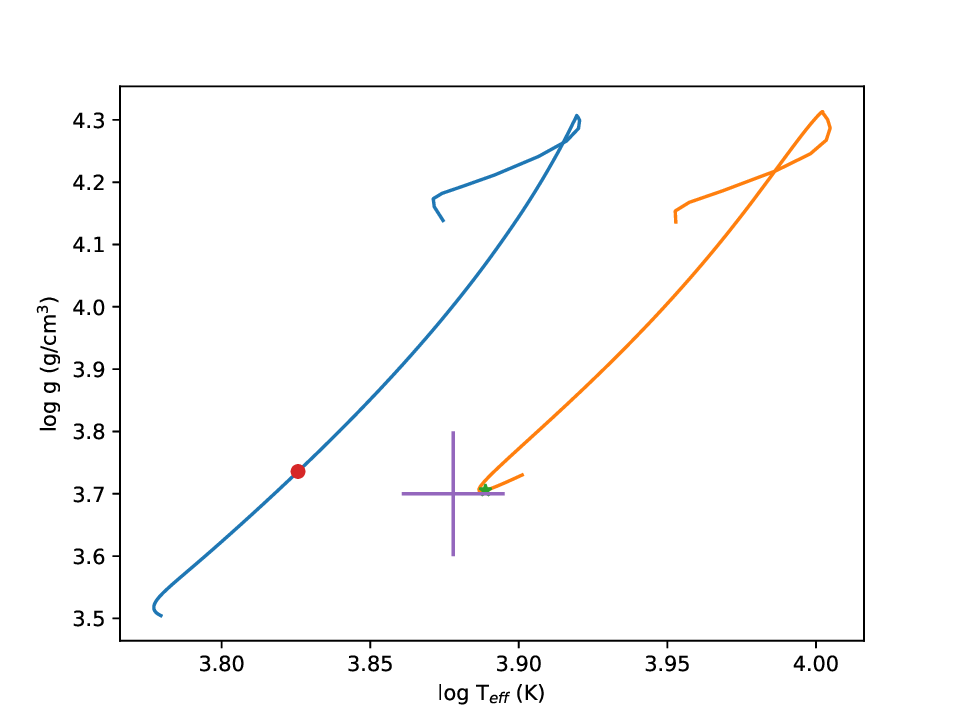}
    \caption{\label{fig:kiel2}The Kiel diagram showing the evolution tracks of the two best-fitting models (see Table~\ref{tab:fitresults}). The observed location of HD~210684 is shown with 1-$\sigma$ errorbars. The location of the best fitting model based on $\ell=1$ (green *) and $\ell=2$ (red $\circ$) are shown.}
\end{figure}

\begin{table}
\setlength{\tabcolsep}{4pt}
\begin{tabular}{cccccccc}
\hline
Mass & Radius & $f_{ov}$ & v$_{ZAMS}$& v$_{eq}$ & Age  & X$_c$ & $\ell$ \\
(M$_{\odot}$)&(R$_{\odot}$)&  & (km s$^{-1}$) & (km s$^{-1}$)& (Gyr) & & \\
\hline
2.3 & 3.52 & 0.0 & 50 & 32.1 & 6.56  & 0.025 & 1\\
1.8 & 3.07 & 0.03 & 40 & 26.6 & 1.45 & 0.223 & 2\\
\hline
\end{tabular}
\caption{\label{tab:fitresults}Properties of the best-fit models.}
\end{table}

We found that the properties of the best fitting models are very similar to 
the observed properties of HD~210684. As shown in Figure \ref{fig:kiel2}, the best fitting model with $\ell = 1$ is within the error bars of the star 
(see Table~\ref{tab:fitresults}).
Nevertheless, the models with $\ell = 2$ result in a better agreement with the found values of the angle between the axis of magnetic dipole and the rotation axis $\beta= 77\degr \pm 3\degr$ (see Subsection~\ref{Bz_measure}), and the equatorial velocity 30.9$\pm$0.6 km s$^{-1}$ (see Subsection~\ref{global}).
The fits with $\ell = 2$ are also more consistent with the observed pulsation, as in our grid, only the coolest models ($\log T_{\rm eff} < 3.85$) are predicted to have positive growth rates (see Appendix~\ref{sim_pulsation}). The model fit assuming $\ell = 1$ pulsations has a negative growth rate, while the $\ell = 2$ model has a positive growth rate.

\begin{figure}
\begin{center}
\includegraphics[width=3.28in,angle=-90]{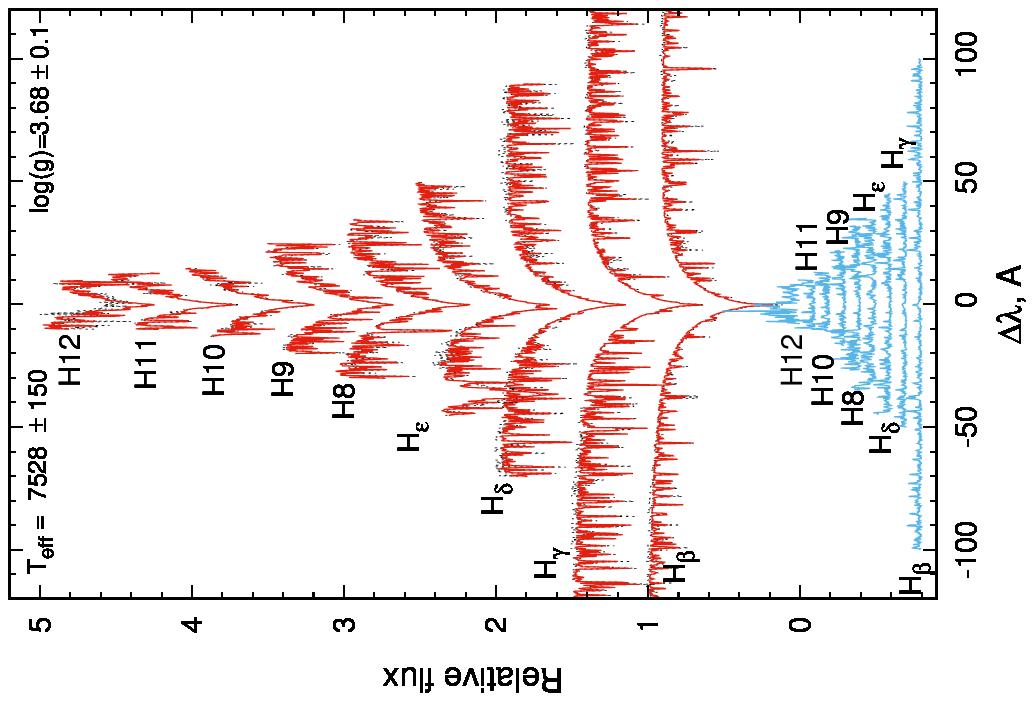} \\
\caption{ An example of fitting the observed Balmer line profiles (black dashed line) of HD~210684 by synthetic spectrum (red line) that corresponds to the model of stellar atmosphere with $T_{\rm eff}$ = 7528$\pm$150~K, $\log{g}$ = 3.68$\pm$0.1, [M/H]= 0, [$\alpha$/H]= 0.33$\pm$0.1, and to the values of v$\sin{i}$ = 16.0$\pm$0.5 km~s$^{-1}$, $v_{\rm r}$= -18.0$\pm$0.5 km~s$^{-1}$ ($\chi^2/\nu$ = 0.3255). Differences between the observed and synthetic profiles are shown at the bottom of this image (blue line). The Balmer line profiles are shifted by 0.5 and the differences are shifted by 0.1 for the sake of visibility. The observed spectrum has been obtained on HJD=2460773.14071 that corresponds to the rotational phase $\varphi$=0.899 (see Table~\ref{tab_Balmer}).}
\label{fig:Balmer_fit}
\end{center}
\end{figure}

\subsection{Fitting of Balmer line profiles}
\label{Balmer}

\begin{table*}[h]
\begin{center}
\caption{Global stellar parameters of HD~210684 derived from the analysis of Balmer line profiles visible in the acquired spectra (see Table~\ref{tab_spectra})
and LSD measurements of the mean longitudinal magnetic field at corresponding rotational phases.
}
\label{tab_Balmer}
\setlength{\tabcolsep}{5pt}
\begin{tabular}{ccccccccc|rr}\hline
 HJD & Rot. & $T_{\rm eff}$, & $\log(g)$ & [$\alpha$/H] &  \multicolumn{2}{c}{$v_{\rm r}$, km s$^{-1}$} & v$\sin{i}$, & $\chi^2/\nu$ & $\langle B_{\rm z} \rangle$, & $\langle N_{\rm z} \rangle$,\\
(2400000+)& phase &  K & & & Balmer & LSD & km s$^{-1}$ &  & G & G \\
\hline
59805.12729 & 0.140 & 7815$\pm$150 & 4.65$\pm$0.10 & 0.36$\pm$0.10 &-17.1$\pm$0.5 &-17.3$\pm$0.1 & 20.0$\pm$0.5 & 0.5385 & 102$\pm$4  & -2$\pm$5 \\ 
59177.70782 & 0.202 & 7784$\pm$150 & 4.47$\pm$0.10 & 0.20$\pm$0.10 &-18.4$\pm$0.5 &-18.6$\pm$0.1 & 19.0$\pm$0.5 & 0.6890 & 155$\pm$27 & 42$\pm$27 \\ 
60282.77044 & 0.252 & 7828$\pm$150 & 4.73$\pm$0.10 & 0.35$\pm$0.10 &-19.2$\pm$0.5 &-19.4$\pm$0.1 & 19.0$\pm$0.5 & 0.5587 & 111$\pm$4  &  3$\pm$4 \\ 
59835.83241 & 0.254 & 7819$\pm$150 & 4.68$\pm$0.10 & 0.34$\pm$0.10 &-19.4$\pm$0.5 &-19.6$\pm$0.1 & 19.4$\pm$0.5 & 0.5379 & 103$\pm$4  &  2$\pm$4 \\ 
60237.84741 & 0.307 & 7784$\pm$150 & 4.65$\pm$0.10 & 0.34$\pm$0.10 &-19.5$\pm$0.5 &-19.4$\pm$0.1 & 19.5$\pm$0.5 & 0.4031 &  84$\pm$4  &  0$\pm$4 \\ 
59459.89670 & 0.394 & 7739$\pm$150 & 4.53$\pm$0.10 & 0.34$\pm$0.10 &-19.6$\pm$0.5 &-19.4$\pm$0.1 & 17.0$\pm$0.5 & 0.3889 &  28$\pm$5  & -1$\pm$5 \\ 
59545.68669 & 0.478 & 7581$\pm$150 & 3.67$\pm$0.10 & 0.36$\pm$0.10 &-19.2$\pm$0.5 &-19.2$\pm$0.1 & 16.3$\pm$0.5 & 0.3664 &  17$\pm$5  &  2$\pm$5 \\ 
59836.97868 & 0.482 & 7570$\pm$150 & 3.66$\pm$0.10 & 0.35$\pm$0.10 &-18.8$\pm$0.5 &-18.8$\pm$0.1 & 16.5$\pm$0.5 & 0.3841 &  10$\pm$4  & -3$\pm$4 \\ 
60093.10957 & 0.485 & 7534$\pm$150 & 3.67$\pm$0.10 & 0.36$\pm$0.10 &-18.7$\pm$0.5 &-18.7$\pm$0.1 & 16.6$\pm$0.5 & 0.3768 &  16$\pm$5  &  2$\pm$5 \\ 
60243.78181 & 0.488 & 7581$\pm$150 & 3.68$\pm$0.10 & 0.37$\pm$0.10 &-18.9$\pm$0.5 &-18.9$\pm$0.1 & 16.4$\pm$0.5 & 0.3693 &  14$\pm$4  &  4$\pm$4 \\ 
59832.00028 & 0.491 & 7573$\pm$150 & 3.68$\pm$0.10 & 0.35$\pm$0.10 &-19.0$\pm$0.5 &-19.0$\pm$0.1 & 16.4$\pm$0.5 & 0.3533 &  20$\pm$4  & -5$\pm$4 \\ 
60238.86927 & 0.510 & 7558$\pm$150 & 3.66$\pm$0.10 & 0.37$\pm$0.10 &-18.1$\pm$0.5 &-18.1$\pm$0.1 & 16.7$\pm$0.5 & 0.3664 &  12$\pm$4  & -0$\pm$4 \\ 
60896.87912 & 0.539 & 7559$\pm$150 & 3.67$\pm$0.10 & 0.38$\pm$0.10 &-18.2$\pm$0.5 &-18.1$\pm$0.1 & 17.3$\pm$0.5 & 0.3909 &   2$\pm$4  &  2$\pm$4 \\ 
61017.75834 & 0.609 & 7574$\pm$150 & 3.65$\pm$0.10 & 0.42$\pm$0.10 &-18.4$\pm$0.5 &-18.4$\pm$0.1 & 18.0$\pm$0.5 & 0.4267 & -31$\pm$4  &  0$\pm$4 \\ 
59837.95679 & 0.677 & 7553$\pm$150 & 3.65$\pm$0.10 & 0.42$\pm$0.10 &-18.7$\pm$0.5 &-18.6$\pm$0.1 & 18.6$\pm$0.5 & 0.4216 & -45$\pm$4  &  2$\pm$4 \\ 
59180.71352 & 0.801 & 7535$\pm$150 & 3.68$\pm$0.10 & 0.38$\pm$0.10 &-19.1$\pm$0.5 &-19.2$\pm$0.1 & 18.2$\pm$0.5 & 0.3670 & -19$\pm$4  &  8$\pm$4 \\ 
59868.84363 & 0.827 & 7537$\pm$150 & 3.67$\pm$0.10 & 0.35$\pm$0.10 &-19.1$\pm$0.5 &-19.3$\pm$0.1 & 17.3$\pm$0.5 & 0.3314 & -17$\pm$4  &  7$\pm$4 \\ 
60280.71959 & 0.844 & 7551$\pm$150 & 3.67$\pm$0.10 & 0.35$\pm$0.10 &-19.6$\pm$0.5 &-19.7$\pm$0.1 & 17.2$\pm$0.5 & 0.3321 & -16$\pm$3  &  3$\pm$3 \\ 
59547.68653 & 0.876 & 7490$\pm$150 & 3.73$\pm$0.10 & 0.31$\pm$0.10 &-18.9$\pm$0.5 &-19.2$\pm$0.1 & 16.7$\pm$0.5 & 0.3512 & -11$\pm$8  & -5$\pm$8 \\ 
59542.68543 & 0.880 & 7511$\pm$150 & 3.70$\pm$0.10 & 0.34$\pm$0.10 &-18.8$\pm$0.5 &-19.0$\pm$0.1 & 16.7$\pm$0.5 & 0.3335 & -10$\pm$4  &  1$\pm$4 \\ 
60773.14071 & 0.899 & 7528$\pm$150 & 3.68$\pm$0.10 & 0.33$\pm$0.10 &-18.0$\pm$0.5 &-18.2$\pm$0.1 & 16.0$\pm$0.5 & 0.3255 &  -9$\pm$4  &  1$\pm$4 \\ 
60316.68966 & 0.998 & 7561$\pm$150 & 3.67$\pm$0.10 & 0.35$\pm$0.10 &-15.9$\pm$0.5 &-16.1$\pm$0.1 & 17.4$\pm$0.5 & 0.3952 &  14$\pm$4  &  1$\pm$4 \\ 
\hline
\end{tabular}
\end{center}
\end{table*}

From the Stokes I spectra, the observed Balmer line profiles and metal lines in the wings were fit with synthetic profiles calculated by \citet{Husser+13} for stellar atmosphere models with different $T_{\rm eff}$, $\log{g}$, and abundance of $\alpha$-elements (O, Ne, Mg, Si, S, Ar, Ca, and Ti) under the condition of solar metallicity for other elements using the code PHOENIX-16 \citep{Hauschildt+97} and an approach similar to \citet{Hauschildt+99}. The code FITSB2 \citep{Napiwotzki+04} was employed to fit the Balmer line profiles and estimate the values of $T_{\rm eff}$, $\log{g}$, abundance of $\alpha$-elements, rotational velocity v$\sin{i}$, and radial velocity $v_{\rm r}$ of the studied target (see Fig.~\ref{fig:Balmer_fit}).
We have used the abundance of $\alpha$-elements as a free fitting parameter assuming that the other chemical species have solar abundance.
HD~210684 possesses a relatively weak magnetic field, and the lowest value of v$\sin{i}$=16.0$\pm$0.5 km s$^{-1}$ derived from the analysis of Balmer line profiles coincides with the rotational phase $\varphi \simeq$ 0.9, where the mean longitudinal magnetic field $\langle B_{\rm z}\rangle$ is close to zero. 
Considering that the absolute value of the measured $\langle B_{\rm z}\rangle$ never exceeds 200~G (see Table~\ref{tab_Balmer}),
we assume here that the weak mean horizontal magnetic field does not contribute substantially to the magnetic broadening.

The best fit values of $T_{\rm eff}$, $\log{g}$, abundance of $\alpha$-elements, rotational velocity v$\sin{i}$, and $v_{\rm r}$ and their associated error-bars based on the fitting of the Balmer line profiles in the Stokes I spectra are shown in Table~\ref{tab_Balmer}  in the columns 3, 4, 5, 6, and 8 respectively, together with the best fit parameter $\chi^2/\nu$ in column 9. The corresponding rotational phases in column 2 have been calculated with respect to the derived rotational period P=5.02188~d (see Subsection~\ref{variability}) taking into account HJDs (column 1) corresponding to the acquisition time of each analyzed spectrum.

The best fits with the lowest values of $\chi^2/\nu$ correspond to the rotational phases $\varphi$=0.49 and 0.90, where $\langle B_{\rm z}\rangle$ is close to zero. The results obtained for the second phase with the lowest $\chi^2/\nu$=0.3255 are shown at Fig.~\ref{fig:Balmer_fit}. It should be also noted that the largest values of $\chi^2/\nu$ are derived for $\varphi$=0.20 and 0.68, where $\langle B_{\rm z}\rangle$ reaches its extrema. 

\begin{figure}
\begin{center}
\includegraphics[width=3.0in,angle=0]{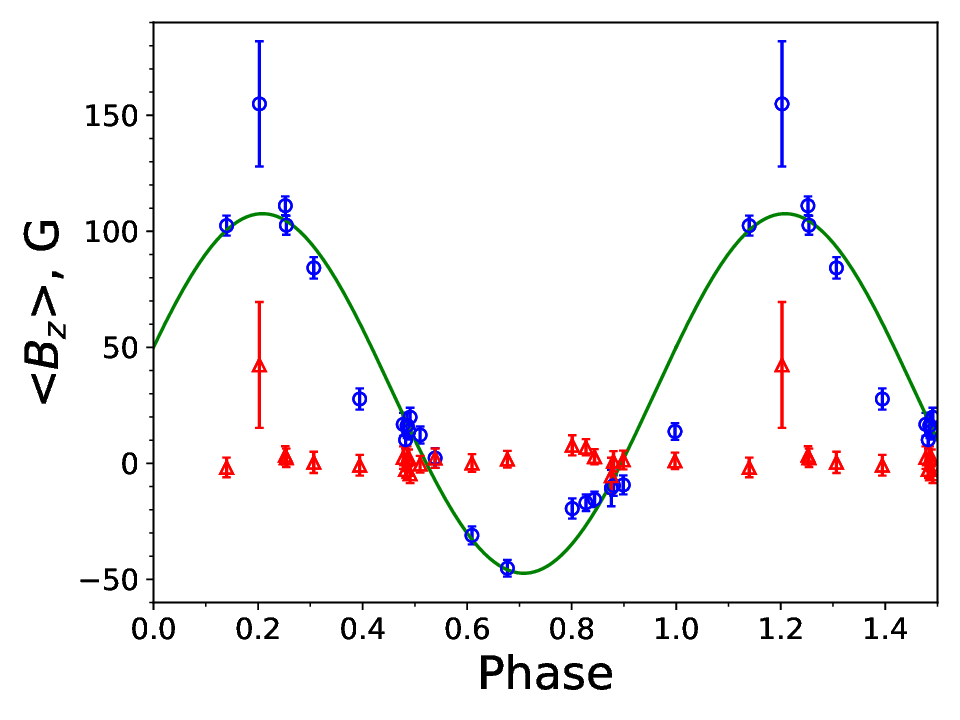}
\caption{Variability of the mean longitudinal magnetic field (blue open circles) with the phase of rotational period P=5.02188~d.
approximated by the simulated $\langle B_{\rm z}\rangle$-curve (green line) assuming model of the centered magnetic dipole.
Red open triangles show level of the possible instrument polarization, $\langle N_{\rm z} \rangle$, inferred from the null profile.
}
\label{fig:Bz_vi}
\end{center}
\end{figure}

\begin{figure}
\begin{center}
\includegraphics[width=2.0in,angle=-90]{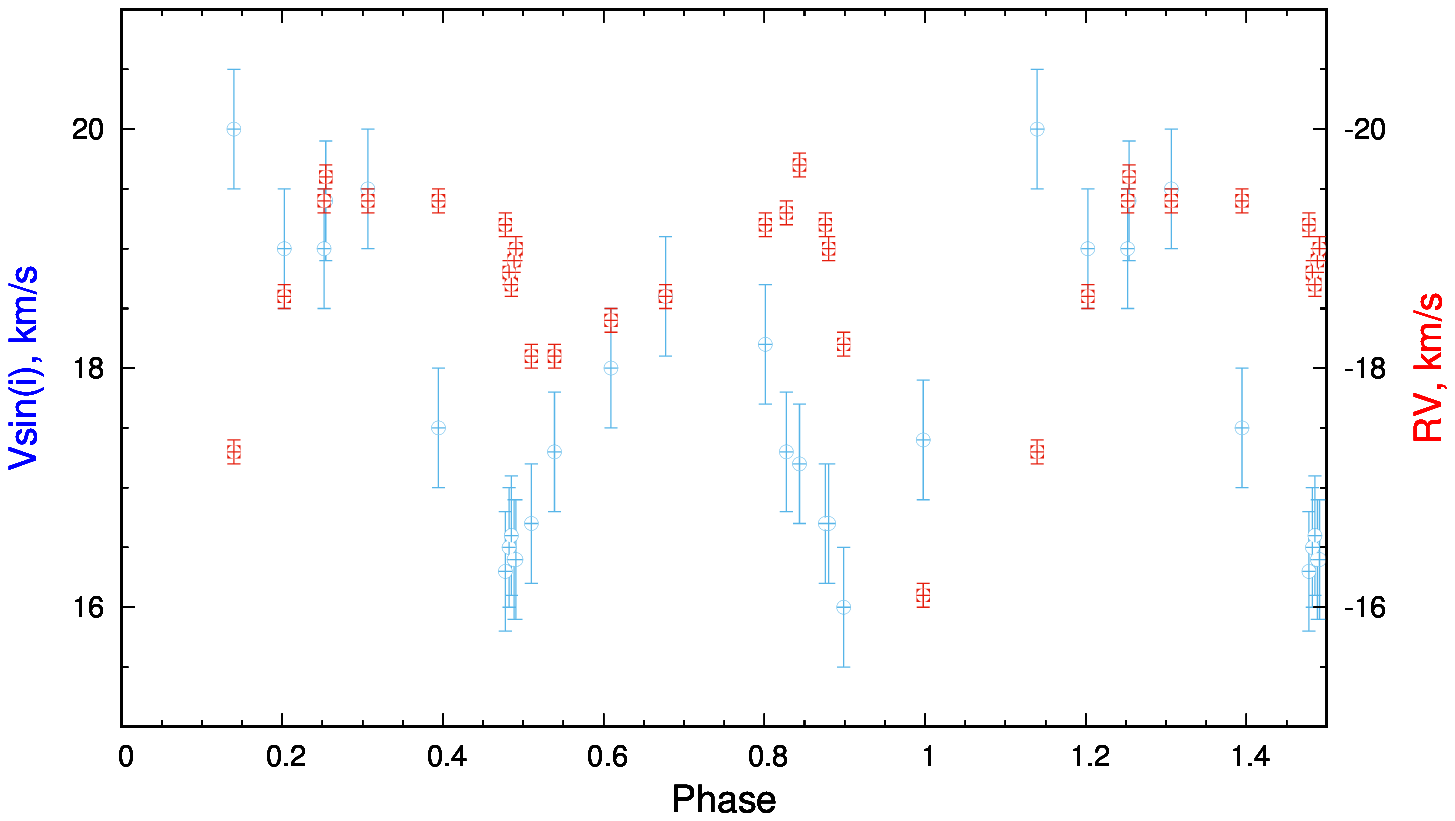}
\caption{
Variability of LSD radial velocity (red open squares) and v$\sin{i}$ (blue open circles) with rotational phase.}
\label{fig:Vr_vi}
\end{center}
\end{figure}

For the rotational phases $\varphi$ = 0.14 -- 0.39, that correspond to the significant positive $\langle B_{\rm z}\rangle$ measurements (see Fig.~\ref{fig:Bz_vi}), the best fit of Balmer line profiles has resulted in a strong overestimation of $T_{\rm eff}$, $\log(g)$ and v$\sin{i}$ with relatively high values of the best fit parameter $\chi^2/\nu$ (see Table~\ref{tab_Balmer}). The aforementioned phases correspond to the first plateau observed at the phased LC (see Fig.~\ref{fig:phased_LC}). It appears that when the positive magnetic pole is most visible in HD~210684 around $\varphi$ = 0.20, one can also see a "hot spot" in its vicinity towards the stellar equator that contributes to our measurements of $T_{\rm eff}$ and $\log(g)$. This hypothesis is supported by the measured radial velocity that changes significantly in this phase interval from -17 km s$^{-1}$ to -19.5 km s$^{-1}$, while the v$\sin{i}$ reaches its maximum values (see Fig.~\ref{fig:Vr_vi}), suggesting that the "hot spot" appears at $\varphi$ = 0.0, crosses the central meridian of the star around $\varphi$= 0.20, and contributes to widening of observed Balmer line profiles.
Nevertheless, to have a clearer picture and explain why the maximum of the derived LC appears at $\varphi$= 0.0 we need to acquire more spectra around this rotational phase.

Considering the high derived values of $\chi^2/\nu$ for the best fits in the phase interval $\varphi$=0.14 -- 0.39 and that our simulation of stellar pulsations in HD~210684 argues in favor of coolest models with $\log T_{\rm eff} < 3.85$ (see Subsection~\ref{puls_results}) we used data from other rotational phases to derive proper values of $T_{\rm eff}$= 7550 $\pm$ 150~K and $\log(g)$=3.7. The best fits obtained for the Balmer line profiles in all phases shows slightly enhanced abundance of $\alpha$-elements [$\alpha$/H]$\sim$0.35 assuming the solar abundance for other chemical species. 



\subsection{Global stellar parameters}
\label{global}

\begin{table} \begin{center}
\setlength{\tabcolsep}{1pt}
\caption{Global stellar parameters known and derived for HD~210684 (TIC~259017938). }
\label{tic_par} \begin{tabular}{lccc}\hline
 Parameter & TIC$^a$ & Gaia$^{b,c}$ & This article  \\
\hline
$T_{\rm eff}$, K      & 7568 $\pm$ 120  & 7610$^{+10}_{-12}$ & 7550 $\pm$ 150 \\
$\log{g}$             & 3.87 $\pm$ 0.08 & 3.811$\pm$0.005 & 3.7 $\pm$ 0.1 \\

 [$\alpha$/H]                &     &  
  & 0.3 $\pm$ 0.1   \\
v$_{\rm r}$, km s$^{-1}$ &  &   & -19.0 $\pm$ 0.5  \\
$L_{\rm \star}$, L$_{\rm \odot}$  & 19.27 $\pm$ 0.64 & & \\ 
$R_{\rm \star}$, R$_{\rm \odot}$  & 2.55 $\pm$ 0.08  & 2.66 $\pm$ 0.13 & 3.07 $\pm$ 0.06 \\ 
$M_{\rm \star}$, M$_{\rm \odot}$  & 1.76 $\pm$ 0.30  & 2.03 $\pm$ 0.10 & 1.8 \\
age, $10^9$ yr &         &   &  1.45 \\
v$\sin{i}$, km s$^{-1}$ &   &            & 16 $\pm$ 0.5  \\
|$\langle B_{\rm z} \rangle$|, G & & & <200 \\
P, days & & & 5.02188$\pm$0.00005 \\
v$_{\rm eq}$, km s$^{-1}$ &   &            & 30.9 $\pm$ 0.6 \\ 
$i$ &   &            & $31\degr \pm 2\degr$    \\ 
$\beta$ &   &          &  $77\degr \pm 3\degr$ \\
\hline \end{tabular} \end{center}
{\it Notes:} $^a$\citet{Stassun+18,Stassun+19}, $^b$DR3: \citet{Fouesneau+22}, 
$^c$DR2: \citet{Kervella+19}.
\end{table}




Using the stellar radius found for HD~210684 by \citet{Kervella+19} and the value of rotational period (see Subsection~\ref{variability}) one can estimate the equatorial velocity from
\begin{equation}
v_{\rm eq}=\frac{50.5927 R_{\rm \star}}{P},
\label{velocity}
\end{equation}
where the stellar radius is measured in R$_{\rm \odot}$, which is recommended by the XXIXth International Astronomical Union (IAU) General Assembly in 2015 to be 695700~km \citep{Prsa+16}, and the period is measured in days. The resulting value for equatorial velocity is 30.9$\pm$0.6 km s$^{-1}$ (see Table~\ref{tic_par}), and, in combination with the most probable value of v$\sin{i}$=16 km~s$^{-1}$ derived from the fitting of Balmer line profiles, it results in an angle $i=31\degr \pm 2\degr$ 
that specifies inclination of the rotational axis to the line of sight.

\citet{Blomme+22} have shown that the template parameters $T_{\rm eff}$, $\log{g}$, and [M/H] \citep{Fouesneau+22} used by CU6 
are not necessarily a good description of the spectral type of the star in he Gaia DR3 archive. 
Nevertheless, the values of $T_{\rm eff}$ and $\log{g}$ derived in this study from the best fit of Balmer line profiles (see Subsection~\ref{Balmer}) are in good accordance with the data found in TESS Imput Catalogue (TIC) \citep{Stassun+18,Stassun+19} and Gaia DR3 archive \citep{Fouesneau+22} considering the error-bars. The effective temperature determined here, $T_{\rm eff}= 7550 \pm 150$~K,  is close to the upper limit found for $T_{\rm eff}$ from the analysis of stellar pulsations. 

Considering that the obtained equatorial velocity 30.9$\pm$0.6 km s$^{-1}$ and the angle $\beta= 77\degr \pm 3\degr$ between the rotation and dipolar axes (see Subsection~\ref{Bz_measure}) better describe the amplitude ratios for the observed pulsation triplet in the best fit model with $\ell=2$ modes (see  Appendix~\ref{mag_OPM}),  
we have derived probable age of HD~210684 to be around 1.45 Gyr (see Table~\ref{tab:fitresults}). Its stellar mass  $M_{\rm \star}$= 1.8~M$_{\rm \odot}$ and radius $R_{\rm \star}$= 3.07~R$_{\rm \odot}$ derived from the simulation of stellar pulsations (see Subsection~\ref{puls_results}) are in good accordance with the data provided by TIC \citep{Stassun+18,Stassun+19}, and Gaia DR3 \citep{Fouesneau+22} archives (see Table~\ref{tic_par}).

\subsection{Magnetic field evaluation}
\label{mag_eval}

\subsubsection{Measurement of the mean longitudinal magnetic field}
\label{Bz_measure}

Normalized Stokes I \& V spectra of HD~210684 have been used to calculate the LSD profile \citep{Donati+97, Kochukhov+2010} and to measure 
$\langle B_{\rm z}\rangle$ with the help of SpecpolFlow software \citep{Erba24, Folsom+25}. This code\footnote{https://folsomcp.github.io/specpolFlow/index.html} was initially developed by Dr. Colin Folsom to analyse 
data provided by spectropolarimeters ESPaDOnS and Neo-NARVAL (or by NARVAL in the past), and is currently updated and supported by the SpecpolFlow team developers 
in the framework of the MOBSTER collaboration.\footnote{https://mobster-collab.com/}

To calculate LSD profiles we created line-masks compiled from the list of atomic data for spectral lines inferred from the VALD3 database \citep{Piskunov+95, Kupka+99, Kupka+00} for our target with $T_{\rm eff}$=7550~K, $\log(g)$=3.7 and [M/H]=0.0 (see Tables~\ref{tab_Balmer},~\ref{tic_par}) assuming microturbulence velocity $v_{\rm mic}$= 2~km~s$^{-1}$ in the spectral domain from 3700\AA\, to 9500\AA. The first mask contains data for 7404 
lines that belong to different chemical species, while the second list of lines with high {\bf Lande} factor (>1.2) includes only 3316 
entries. Considering that the derived v$\sin{i}$= 16.0~km~ s$^{-1}$ we adopted LSD width of 20~km~ s$^{-1}$ to calculate the mean longitudinal magnetic field. The use of two different line-masks results in almost the same $\langle B_{\rm z} \rangle$ values, but we report here only the data obtained with the line-mask containing solely lines with the high {\bf Lande} factor because the final error-bars appear to be more realistic as compared to the measurement of polarized signal $\langle N_{\rm z} \rangle$ from the null profiles (see Fig.~\ref{fig:Bz_vi}). Corresponding radial velocities were derived {\bf by fitting a Gaussian function to LSD Stokes I profiles} (see column 7 for LSD $\rm v_{r}$ in Table~\ref{tab_Balmer}). For each studied rotational phase, they are in a good accordance with the radial velocities derived from the fitting of Balmer line profiles.

Figure~\ref{fig:Bz_vi} presents the obtained $\langle B_{\rm z}\rangle$ values that show clear variability with the rotational period P=5.02188~d derived from the \textit{TESS} photometry. 
The mean longitudinal magnetic field reaches its maximum at the rotational phase $\varphi$= 0.21 which is shifted with respect to the maximum on the phased LC at $\varphi$= 0.0 (see Fig.~\ref{fig:phased_LC}). Unfortunately, the measured $\langle B_{\rm z}\rangle$ maximum has the largest error-bar because it corresponds to the spectrum with the lowest SNR (see Table~\ref{tab_spectra}). The minimum of $\langle B_{\rm z}\rangle$ with significant field is found around the phase $\varphi \sim$ 0.68 that corresponds to the second plateau on the phased LC observed after the minimum (see Fig.~\ref{fig:phased_LC}). The employed LSD method provides quite small values of the significant mean longitudinal magnetic field, which is less than 200~G in its maximum.

In the case of centered magnetic dipole 
the oblique magnetic rotator (OMR) model provides the following expression 
\citep[see][for more detail]{Babcock58,Mathys88,Bagnulo+96,Khalack02}:
\begin{equation}
\langle B_{\rm z}\rangle \simeq B_{\rm p}(\cos{\beta}\cos{i}+\sin{\beta}\sin{i}\cos{(\varphi+\varphi_{0}})),
\label{Bz_approximation}
\end{equation}
to approximate the measured $\langle B_{\rm z}\rangle$ variability with a theoretical curve (see green line in Fig.~\ref{fig:Bz_vi}).
Here, $B_{\rm p}$ specifies intensity of the surface magnetic field at the magnetic pole, 
and $\varphi_{0}$ is the phase shift. 
We can see that the positive magnetic pole is most visible at the rotational phase $\varphi \sim$ 0.21, while the negative magnetic pole appears around $\varphi \sim$ 0.68. {\bf By} fitting (using the \textit{curve\_fit} routine from python's scipy.optimize\footnote{https://docs.scipy.org/doc/scipy/reference/optimize.html} library) {\bf Eq.~\ref{Bz_approximation} to the} mean longitudinal magnetic field measurements phased with the rotational period we derived the angle between {\bf the} dipole and rotation axes
$\beta= 77\degr \pm 3\degr$
for the known inclination of the rotation axis to the line of sight $i=31.2\degr \pm 1.8\degr$. The fact that the derived approximation does not fit $\langle B_{\rm z}\rangle$ measurements {\bf well at all} indicates that the real surface magnetic field structure is different from the dipolar configuration, and higher multipoles \citep{Bagnulo+96} or a decentered magnetic dipole \citep{Khalack02,Khalack05} should be considered.

\subsection{Rotational variability of the observed LSD line profiles}
\label{LSD_var}

\begin{figure}
\begin{center}
\begin{tabular}{cccc}
\includegraphics[width=3.2in,angle=-90]{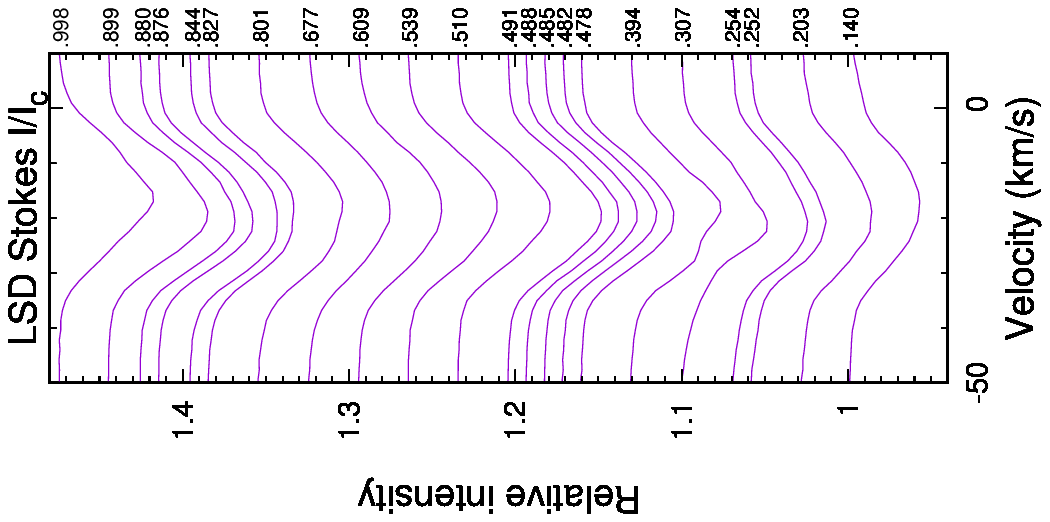} &
\includegraphics[width=3.2in,angle=-90]{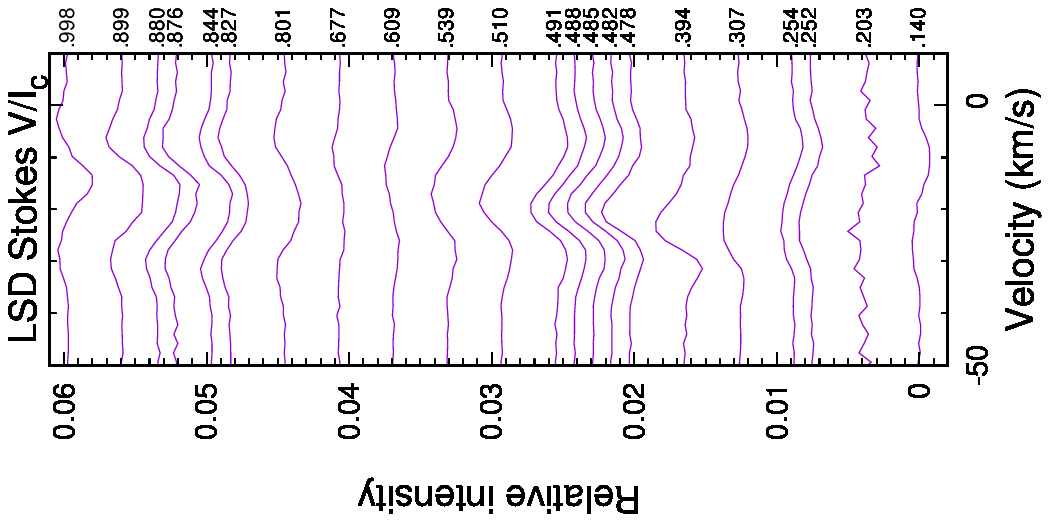} \\
\end{tabular}
\caption{Variability of the LSD Stokes I (left) and V (right) profiles with rotational phase. LSD profiles are shifted vertically with the rotational phase for the sake of visibility, but those shifts remain small for the same of similar phases. }
\label{fig:LSD_prof}
\end{center}
\end{figure}

The LSD Stokes I \& V profiles simulated for each spectropolarimetric {\bf spectrum} of HD~210684 (see Subsection~\ref{Bz_measure}) were phased with the rotational period to study their variability that could contain information about a horizontal abundance stratification of different chemical species. {\bf Despite} the derived $\langle B_{\rm z} \rangle$ {\bf being} very weak, the simulated LSD Stokes V \& I profiles show clear variability with the rotational phase (see Fig.~\ref{fig:LSD_prof}). 


The simulated LSD Stokes I profiles show significant asymmetry at the next intervals of rotational phases $\varphi$= 0.25 -- 0.39 and 0.80 -- 1.00, suggesting that {\bf a} horizontal stratification of chemical abundance (and respectively of surface temperature) is most pronounced at these phases. The strongest asymmetry of the LSD Stokes I profile is observed at $\varphi$= 0.307 and 0.394, where the radial velocity remains constant, v$\sin{i}$ and $\langle B_{\rm z} \rangle$ drop, and the phased LC also decreases after passing the first plateau (see Figs.~\ref{fig:phased_LC},~\ref{fig:Bz_vi} and \ref{fig:Vr_vi}). The observed variability and asymmetry of the LSD Stokes I profiles indicates a presence of horizontal abundance stratification in stellar atmosphere of {\bf HD~210684}.


\section{Discussion}
\label{discussion}

A detailed analysis of the TESS light curve for HD~210684 in sectors 56 \& 83 revealed a triplet of high-overtone pulsations 
around the frequency $\nu_{\rm p}$=116.825~$d^{-1}$ which confirms its classification as a roAp type variable \citep{Holdsworth+24}. We have {\bf found} a signal at lower frequencies  that corresponds to the rotational modulation of the light curve with the period P=5.02188$\pm$0.00005~d.
When the estimated error bar on this rotational period is taken account, it 
appears to be slightly lower than the one reported by \citet{Balona22}. 
In our asteroseismic models, the driving region for these pulsations occurs quite close to the surface ($>0.85R_*$ in both cases). Assuming that the observed triplet at 116.825~$d^{-1}$ is rotationally split by $\Delta\nu$=0.19913~$d^{-1}$ (P= 5.0219$\pm$0.0009~d) in the driving region,
we see no evidence of differential rotation in this star.

We compared the observed triplet with model frequencies calculated using MESA and GYRE. We do not provide
tight constraints on the star. Overall, we find that models with high overshoot and low mass are ruled out by comparison with the observed rotational splitting.  There does not seem to be a strong preference for a particular overshoot parameter based on our model grid. Overall, we find that very few of the models in our grid are predicted to have driven frequencies in the range of observed frequency triplet.  Only the most evolved, coolest models show values of $\eta > 0$ for both $\ell=1$ and $\ell=2$ modes. This suggests the star has evolved quite far along the main sequence.
The best fit model parameters depend on whether the triplet is assumed to be $\ell = 1$ or $\ell =2$. In the $\ell =1 $ case, the mass is higher and the overshoot is lower than in the $\ell = 2$ case. 
Both of the best fit models are in good agreement with the observed $\log T_{eff}$ and $\log g$ of the star (see Fig.~\ref{fig:kiel2}). However, in our best fit model the $\ell=1$ frequencies are not predicted to be excited. In addition, the modes with $\ell =2$ result in the angle between axes of magnetic dipole and stellar rotation of $\beta \sim 90\degr$\ (see Appendix~\ref{mag_OPM})
closer to the observed value of $\beta= 77\degr \pm 3\degr$ derived under the assumptions of the OMR model from the observed variability. 
As such, we believe that preference should be given to models with
$\ell =2$ modes. However, it should be noted that our simulations were carried out without considering {\bf the effects of any} 
magnetic field. 

Our estimates of the global stellar parameters derived from 
fitting Balmer line profiles observed in ESPaDOnS spectra
are in good agreement 
with the data inferred from TIC \citep{Stassun+18,Stassun+19} 
and DR3 \citep{Fouesneau+22} archives (see Table~\ref{tic_par}). Using the stellar radius provided for HD~210684 by \citet{Kervella+19}, its rotational period and v$\sin{i}$ derived in this study we estimated the equatorial velocity (see Eq.~\ref{velocity}) and the angle of inclination of the rotation axis to the line of sight $i=31\degr \pm 2\degr$. 
{\bf Using} the model with $\ell =2$ modes one can predict 
that HD~210684 continues its evolution on the main sequence (see Fig.~\ref{fig:kiel2}) having {\bf an age of} 1.45 Gyr.

Application of the LSD approach to the normalized Stokes I \& V spectra resulted in detection of the mean longitudinal magnetic field
that varies with the derived period of stellar rotation. 
It reaches its maximum 
at $\varphi$= 0.20, {\bf which} corresponds to the beginning of a plateau on the phased LC (see Subsection~\ref{Bz_measure}). The derived $\langle B_{\rm z}\rangle$ measurements are quite small (< 200~G), but significant at the rotational phases when positive or negative magnetic poles are most visible. Variability of the $\langle B_{\rm z}\rangle$ measurements phased with the rotational period appears to be different from the curve predicted by the OMR model of a centered magnetic dipole (see Fig.~\ref{fig:Bz_vi})
suggesting that configuration of the surface magnetic field 
is more complex than a dipole.

Considering that HD~210684 is a probable member of a binary system \citep{Kervella+19}, we performed a preliminary analysis of the spectra and found no obvious spectral lines that may belong to a secondary component. 
Given the mass of the secondary proposed by \citet{Kervella+19}, the contribution of secondary component to the acquired spectra is expected to be negligible, which is consistent with our results.
The absence of a second component in the spectrum suggests that all detected variability of the simulated LSD I profiles (see Fig.~\ref{fig:LSD_prof}) is caused by inhomogeneous horizontal distribution of elements abundance in stellar atmosphere of the primary component, which is a weakly magnetic CP star and roAp variable.
Considering that the best fit of Balmer line profiles in the spectra acquired for $\varphi$= 0.14 -- 0.39 results in significantly higher values of $T_{\rm eff}$, $\log(g)$ and v$\sin{i}$, we assume that a "hot spot" visible at the aforementioned rotational phases may contribute to these effects. This hypothesis is supported by strong asymmetry of the simulated LSD I profile at $\varphi$= 0.307 and 0.394 suggesting a link between the location of "hot spot" and patches of enhanced (or depleted) abundance for several chemical species.

As we do not have good coverage of all rotational phases with spectropolarimetric observations, additional high-SNR spectra of HD~210684 are required to carry out detailed mapping of the horizontal abundance stratification for different chemical species and reconstruction of the surface magnetic field configuration.
The combination of a weak ($\langle B_{\rm z}\rangle$ < 200G) and significantly non-dipolar magnetic field, the presence of an area with hotter local surface temperature, a complex structure of horizontal abundance stratification, and relatively strong high-overtone pulsations of roAp type 
make this a fascinating target for future study.
A comprehensive study of its nature will provide observational constraints for theoretical simulation of magnetic field generation in magnetic CP stars \citep{Schleicher+23}, its impact on the horizontal and vertical abundance stratification of chemical elements \citep{Alecian+Stift21, Alecian23} and for GYRE models used to calculate driven frequencies of high-overtone pulsations in roAp stars.


\begin{acknowledgements}
\label{acknow}

V.K. acknowledges support from 
Mitacs, ACFAS, and is thankful to the Facult\'{e} des \'{E}tudes Sup\'{e}rieures et de la Recherch and to the Facult\'{e} des Sciences de l'Universit\'{e} de Moncton for financial support of this research. CCL acknowledges support from the Natural Sciences and Engineering Research Council of Canada (NSERC).
The authors are grateful to Dr. Kurtz, Dr. Holdsworth and Dr. Kochukhov for useful advices that helped to improve significantly this article.
The analysed spectra have been obtained at the Canada-France-Hawai`i Telescope (CFHT) which is operated by the National Research Council of Canada, the Institut National des Sciences de l'Univers of the Centre National de la Recherche Scientifique of France, and the University of Hawai`i. {\bf CFHT is located on Maunakea on Hawai`i Island, a mountain of considerable cultural, natural, and ecological significance. Maunakea is a sacred site to Native Hawaiians, also known as K\={a}naka `\={O}iwi. Quality observations are made possible by relentless effort of the entire staff at Canada-France-Hawai`i Telescope.} 
Authors thank to Oleksandr Kobzar for sharing his spectra for HD~210684 and to the \textit{TESS} and TASC/TASOC teams for their support of the present work.
This paper includes data collected by the \textit{TESS} mission. Funding for the \textit{TESS} mission is provided by the NASA Explorer Program. This research has made use of the SIMBAD database, operated at CDS, Strasbourg, France. Some of the data presented in this paper were obtained from the Mikulski Archive for Space Telescopes (MAST) and Gaia DR3 catalogue.
\end{acknowledgements}

\bibliographystyle{aa}
\bibliography{AllDelta_roAp_New_HD159541}

\begin{appendix}




\onecolumn
\section{Oblique pulsator model}
\label{mag_OPM}

The DFT calculated with the code \textit{Period04} for the region of high frequencies (see Fig.~\ref{fig:puls}) also provides the amplitudes of high-overtone pulsations that can be used in the frame of oblique pulsator model (OPM) \citep{Kurtz82} to estimate the angle $\beta$ between the axis of magnetic dipole and the rotation axis. 
Following the frequency analysis performed by \citet{Handler+06} for HD~99563 one can apply the OPM that provides a relation for pulsation modes with $\ell$=1 \citep{Shibahashi86}:

\begin{equation}
\frac{A^{(1)}_{+1}-A^{(1)}_{-1}}{A^{(1)}_0}=\sin{\beta} \sin{i},
\label{apl_ratio_1}
\end{equation}
where $A^{\rm (1)}_{\rm 0}$ is the amplitude of the central frequency in the high-overtone triplet, $A^{\rm (1)}_{\rm +1}$ and $A^{\rm (1)}_{\rm -1}$ are, respectively, the amplitudes of the higher and lower rotational
side-lobes, and $i$ specifies angle between the line of sight and the rotational axis of the star. For the derived amplitudes of high-overtone pulsations 
equation~(\ref{apl_ratio_1}) results in $\sin{\beta} \sin{i}= -0.07\pm0.07$ 
Considering the known angle $i=31\degr \pm 2\degr$ (see Table~\ref{tic_par})
one can derive that the angle $\beta$ is relatively small or close to zero
( $\beta\simeq {-8\degr}^{+8\degr}_{-9\degr}$).

If the observed triplet of high-overtone pulsations consists of modes with $\ell$=2 
one should employ the following relation \citep{Kurtz92, Shi+21}:
\begin{equation}
\frac{A^{(2)}_{+1}+A^{(2)}_{-1}}{A^{(2)}_0}= \frac{12 \sin{\beta}\cos{\beta} \sin{i}\cos{i}}{(3\;cos^2{\beta}-1)(3\;sin^2{i}-1)},
\label{apl_ratio_2}
\end{equation}
where $A^{\rm (2)}_{\rm 0}$ is the amplitude of the central frequency, $A^{\rm (2)}_{\rm +1}$ and $A^{\rm (2)}_{\rm -1}$ are, respectively, the amplitudes of the higher and lower rotational
side-lobes. For the known angle $i$ this relation can be transformed into a quadratic equation with respect to $\cos{(2\beta)}$ and solved analytically:
\begin{equation}
cos{(2\beta)} = \frac{-3\;b^2 \pm 4\;a \sqrt{2\;b^2+a^2}}{9\;b^2+4\;a^2},
\label{2beta}
\end{equation}
where
\[ a=6\;sin{i}\cos{i}, \;\;\; b= \frac{A^{(2)}_{+1}+A^{(2)}_{-1}}{A^{(2)}_0}(3\;sin^2{i}-1). \]
For the derived value of angle $i$ 
the equation~(\ref{2beta}) provides 
$\beta \sim 0\degr$ (for positive sign in front of the square root) and $\beta \sim 90\degr$ (for negative sign).
Clearly the error bars are large in this case, and no constraints can be placed on the magnetic field orientation.

\section{Simulation of stellar pulsations}
\label{sim_pulsation}

\begin{figure*}[h]
\begin{center}
\includegraphics[width=3.5in]{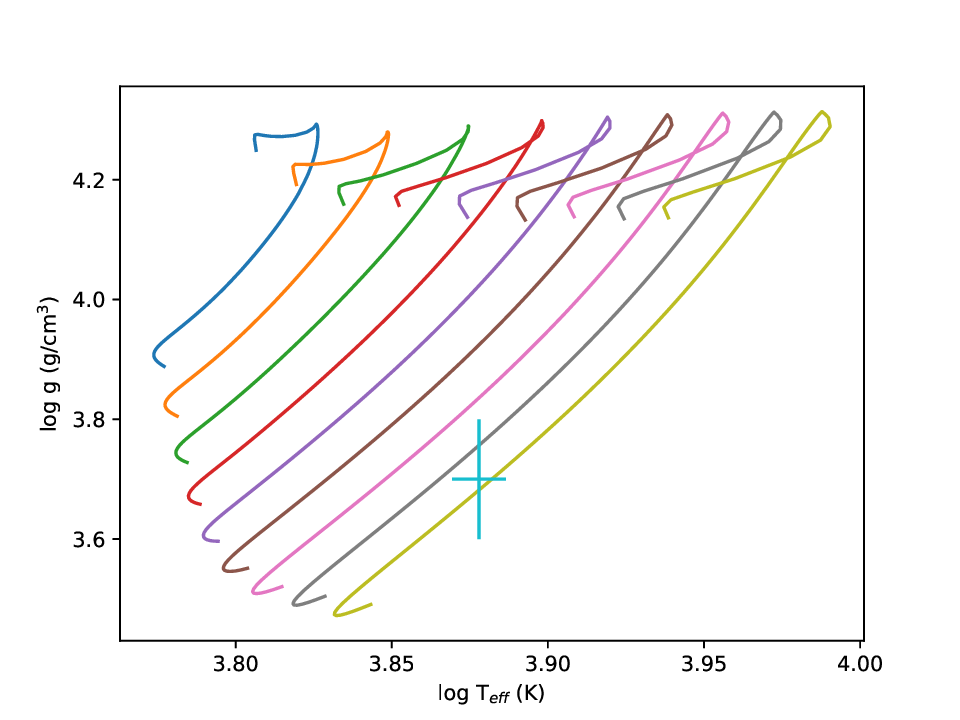}
\end{center}
\caption{Evolution tracks for 1.4 - 2.2~M$_{\odot}$ models in a Kiel diagram.  All models in this figure have a rotation rate of 50 km s$^{-1}$ on the ZAMS, and a convective overshoot parameter of $\alpha = 0.02$. The position of HD~210684 is shown, with the errorbars indicating 1$\sigma$ uncertainties on the observed effective temperature and surface gravity.  }
\label{fig:HR}
\end{figure*}

We have calculated a grid of stellar structure and evolution models using MESA version 12778 \citep{Paxton+11, Paxton+13, Paxton+15, Paxton+18, Paxton+19}.  The grid covers main sequence models from 1.4 to 2.2 M$_{\odot}$ in increments of 0.1 $M_{\odot}$, including the effects of convective overshoot and rotation.  Rotation rates varied from 30 km s$^{-1}$ to 100 km s$^{-1}$ on the ZAMS based on the rotational velocity derived from spectroscopy. Convective mixing was calculated using the mixing length theory \citep{Bohm-Vitense54} with a mixing length $\alpha = 2.0$. Convective overshoot was included above and below convective zones using the exponential formalism \citep{Herwig2000} with the overshoot parameter ranging from 0 to 0.04. As discussed below, spectral fitting to the Balmer line profiles provided [M/H]=0 with slight overabundance of $\alpha$-elements [$\alpha$/H]= $0.3 \pm 0.1$, consistent with a metallicity of Z = 0.02, which we have adopted for our model grid. Evolution tracks are shown for $f_{ov}$ = 0.02 and $v_{ZAMS} = 50$ km s$^{-1}$ in Figure~\ref{fig:HR}.  The observed position of HD~210684 is marked and the errorbars show the 1$\sigma$ uncertainties on the parameters based on the fits to Balmer lines.

\begin{figure*}[h]
\includegraphics[width=0.5\textwidth]{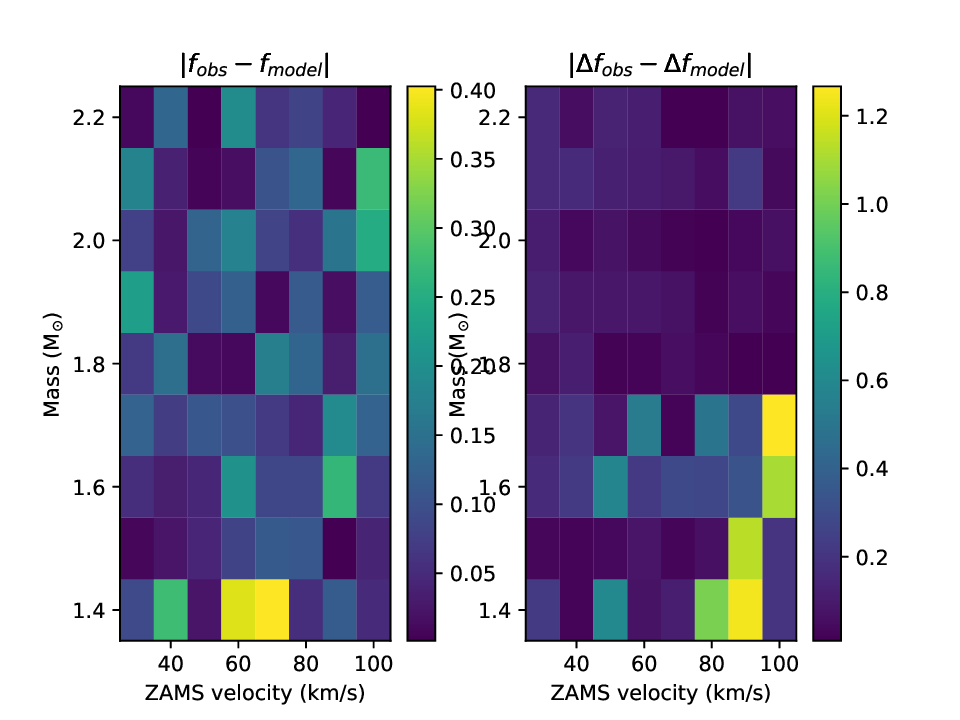}
\caption{\label{fig:fov_fit}The difference between the model $\ell = 2$ frequency and observed central frequency (left panel) and the observed and model splitting (right panel) for models with overshoot parameter $f_{ov}$=0.04. Each point represents an evolution track, and the model that best matches the central frequency has been taken along each track. As expected, there are many possible values based on the central frequency alone, making it hard to constrain this star. The splitting values do rule out some regions of parameter space. }
    \end{figure*}

\begin{figure*}[h]
    \includegraphics[width=0.5\textwidth]{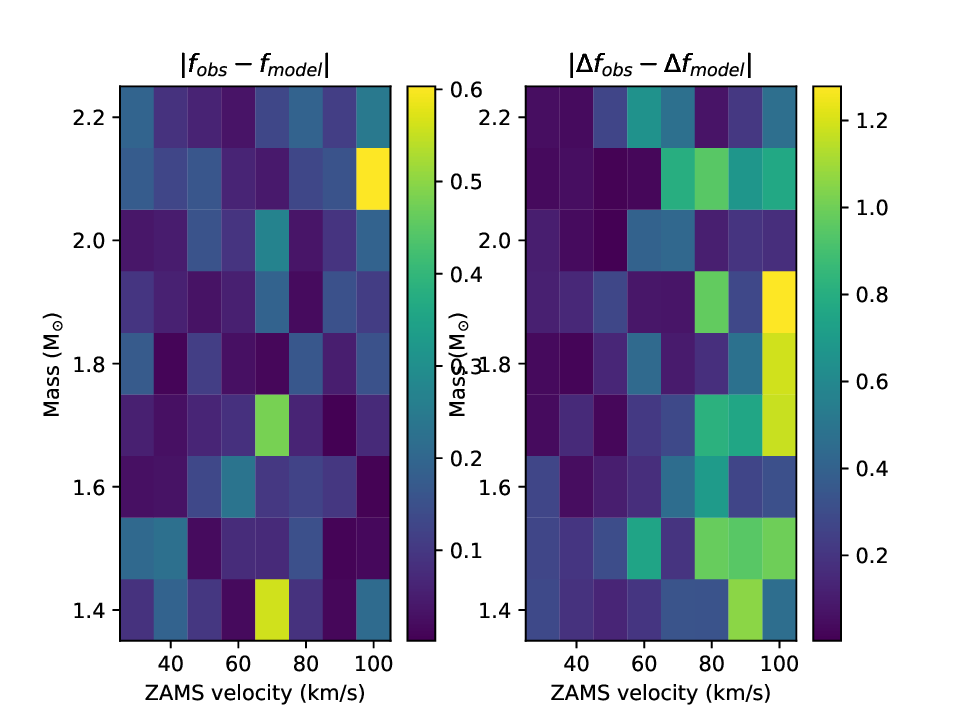}
    \caption{\label{fig:fov_01} The same as for Figure \ref{fig:fov_fit} for models with $f_{ov} = 0.01$. The region excluded by poor match to the observed splitting becomes much larger at lower overshoot.}
\end{figure*}

We saved detailed models every 10 time-steps, and used these to calculate linear non-adiabatic pulsation frequencies with GYRE \citep{Townsend+Teitler13, Townsend+18}.  Since only one frequency triplet is observed, we calculated theoretical frequencies over a small range, from 115~d$^{-1}$ to 118~d$^{-1}$ for $\ell$ = 1 and 2, with $m$ = -1, 0, or +1.
At each model, we identified the frequency triplet with a central frequency closest to the observed frequency, $\nu$ = 116.8250~d$^{-1}$. Using this triplet, we calculated the absolute value of the difference from the central frequency.  The model splitting was calculated using the average of the $(m=1)-(m=0)$ and $(m=0)-(m=-1)$ splitting. The distribution of these two fit parameters are shown for $f_{ov} = 0.04$ in Figure \ref{fig:fov_fit}.

As shown in Figure \ref{fig:fov_fit}, with only a single triplet, we can place very limited constraints on the stellar parameters.  Matching the splitting does rule out some areas of the parameter space, as can be seen in the right panel of Figure \ref{fig:fov_fit}.  Models with high rotation rates and low mass have splitting that is a poor match to the observations.  At lower overshoot parameters, this region becomes much larger, as shown in Figure \ref{fig:fov_01}.

Overall, our models favour the higher end of the mass range and lower initial rotation velocities, regardless of overshoot. The vast majority of our fits correspond to evolved main sequence models, and those that are fit to young models using the central frequency tend to be excluded by the fits to the splitting.

\begin{figure*}[h]
    \includegraphics[width=0.5\textwidth]{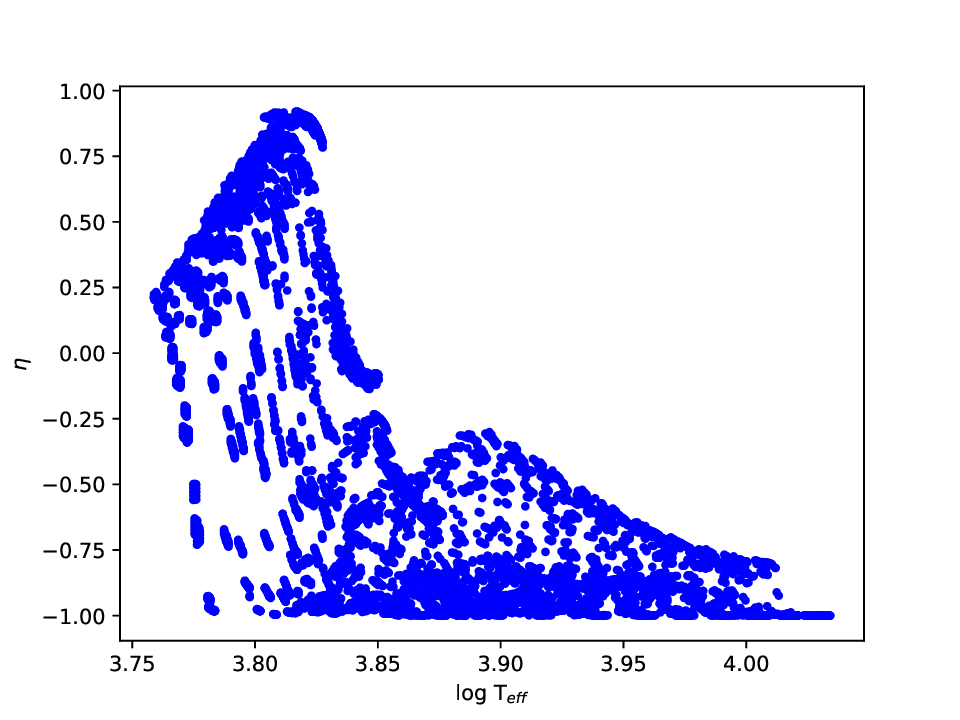}
    \caption{\label{fig:excitation}The predicted growth rates for all $m=0$ modes in our model grid.  Only some of the models with $\log T_{\rm eff} < 3.85$ show positive growth rates.}
\end{figure*}

Finally, we looked at the predicted excitation of the modes in our models, and found that neither of the best fitting models predict excited frequencies in this region.  Both models have $\eta \approx -0.9$, indicating the frequencies should be strongly damped 
throughout their evolution. We looked at the growth rates of all $m=0$ modes in our grid, and found that the growth rates are only predicted to be positive for some of the coolest models ($\log T_{\rm eff} < 3.85$), as shown in Figure \ref{fig:excitation}. In general, the models that show excitation correspond to the most evolved models in our grid.

\end{appendix}
\end{document}